\documentstyle[12pt]{article}
\begin{document}

{\large 
\centerline{\bf  Recent Advances in Hamiltonian
Lattice Gauge Theory \footnote{Review talk at Beijing workshop on
Experiemental and Theoretical Study of Glueballs}} 
}

\vskip 0.3cm

\centerline{\bf Shuo-Hong Guo, ~~ Xiang-Qian Luo
\footnote{E-mail address: STSLXQ@ZSULINK.ZSU.EDU.CN}}

\vskip 0.3cm

\centerline{\it CCAST (World Laboratory), Beijing 100080,  China}

\centerline{\it Department of Physics, Zhongshan University,
Guangzhou 510275, China
\footnote{Mailing address}}

\vskip 0.3cm

\centerline{\bf Abstract}

We review the recent advances in the Hamiltonian formulation of lattice
gauge theory for approaching the continuum physics. In particular, vacuum
wave function and glueball spectrum calculations by coupled cluster method
with truncation scheme preserving the continuum behavior are described.

\vskip 1cm

\section{Introduction}

Lattice gauge theory (LGT) has now developed into one of the most
informative method for non-perturbative aspects of strong interactions. It
is a fundamental theory based on first principle of strong interactions. In
1974, Wilson proposed the Lagrangian formulation of LGT \cite{Wilson} where
the gauge field theory is discretized on the D-dimensional space-time
lattice. In 1975, Kogut and Susskind derived a lattice Hamiltonian \cite{KS}
in which only the (D-1)-dimensional space is discretized while the time
variable remains continuous. Theoretically, these two approaches are
equivalent in the continuum limit.

For simplicity, we consider only the case of pure gauge fields. In the
continuum, the action is

\begin{eqnarray}
S=-{1 \over 4} \int d^D x ~ ~ {\cal F}_{\mu \nu}^{\alpha} {\cal F}_{\mu \nu}^{\alpha}.
\end{eqnarray}
Continuum quantum field theory with such an action suffers from divergence
problem, which has to be regularized and renormalized. In perturbative QCD,
the theory is regularized at high momenta, and renormalization group
equation gives the effective coupling constant ${\bar g}$ as a function of
the momentum scale $\mu $, i.e. ${\bar g}={\bar g}(\mu ).$ For SU(N) group, $%
{\bar g}$ goes to zero when $\mu \to \infty $, so that the high energy
strong interactions can be well described by perturbative QCD. For low
energy phenomena, such as quark confinement and glueball spectrum, where ${%
\bar g}$ becomes large, non-perturbative methods must be employed.

LGT uses the space-time discretization to regularize the continuum theory.
On the lattice, the gauge field $A$ is replaced by the link variable $%
U_l(x_0)=exp(ig\int_{x_0}^{x_0+a{\vec l}}dx_lA_l)$, where $a$ is the lattice
spacing. Renormalization group equation predicts the so-called scaling
relation $g=g(a)$. $g$ goes to $0$ as $a\to 0$. Therefore, the continuum
limit corresponds to the weak coupling limit of the lattice theory. In this
limit, the evolution of $g(a)$ is the same as the evolution of ${\bar g}(\mu
)$ in the continuum theory, with $a\approx \mu ^{-1}$. For a 4D SU(N)
theory, the scaling relation is

\begin{eqnarray}
\Lambda_L a = g^{-{b_1 \over 2b_{0}^{2}}} e^{-{1 \over 2b_{0} g^{2}}} (1+c_1 g^2 + ...) .
\end{eqnarray}
The constants $b_0$, $b_1$ and $c_i$ can be calculated by continuum
perturbation theory. $\Lambda _L$ is the lattice scale parameter with
dimension of mass, and it can be related to the continuum scale parameter
such as $\Lambda _{\bar {MS}}$. When $a$ varies, the lattice coupling
constant $g$ must vary accordingly in order to give the same physical
results for continuum physics. On the lattice, one calculates dimensionless
quantities such as $am$, with $m$ some physical mass. In the weak coupling
region, $am$ should scale as $const.\times \Lambda _La(g)$. Correct scaling
behaviors is crucial for extracting physical results from lattice
calculations.

The last two decades has seen a lot of progress in Monte Carlo (MC)
simulations of LGT in the Lagrangian formulation, while analytic and
numerical investigations of Hamiltonian formulation were not so active.
Nowadays, MC simulation of lattice QCD is able to reach $\beta =6/g^2\approx
6.4$, but it is unfortunately not weak enough for the scaling behavior to be
satisfied, and improvement is a pressing task. In recent years there have
been some remarkable advances in the Hamiltonian formulation for approaching
the continuum physics. It now seems that the Hamiltonian formulation
deserves much attention.

\section{Hamiltonian versus Lagrangian Formulation}

The Lagrangian formulation starts from the path integral $Z=\int
[dU_l]e^{-S[\{U_l\}]}$. Wilson \cite{Wilson} proposed an action

\begin{eqnarray}
S={1 \over g^2} \sum_p Tr(U_{p} +U_{p}^{\dagger}-2),
\end{eqnarray}
where $U_p$ is the ordered product of $U$ around an elementary plaquette.
The mass spectrum can be obtained by computing the correlation function $%
C(t)=\langle \Phi (t)\Phi (0)\rangle $, where $\Phi (t)$ is some appropriate
operator with nonzero projection onto the lowest excited state. Then

\begin{eqnarray*}
C(t) = Z^{-1} \int [d U_l] e^{-S[\{ U_l \}]} \Phi (t) \Phi (0) = Tr[e^{Ht} \Phi (0) e^{-Ht} \Phi (0)]
\end{eqnarray*}
\begin{eqnarray}
= \sum_n \vert \langle \vert \Phi (0) \vert n \rangle \vert^2 e^{-(\epsilon_n-\epsilon_0)t} 
\rightarrow^{t \to \infty}  e^{-(\epsilon_1-\epsilon_0)t} .
\label{COR}
\end{eqnarray}
For (\ref{COR}) to hold at not too large $t$, $\Phi $ must have sufficiently
large projection onto the lowest excited state. Furthermore, statistics must
be sufficient and finite size effects must be under control. Despite
existing problems, the Lagrangian formulation simulated by MC algorithm is
still the most efficient way for obtaining low energy hadron mass spectrum.

In contrast, the Hamiltonian formulation starts from the lattice Hamiltonian 
\cite{KS}

\begin{eqnarray}
H={g^2 \over 2a} \sum_{l} 
 E_l^{\alpha} E_l^{\alpha}
 -
{1 \over g^2 a} \sum_{p} 
Tr(U_{p} +U_{p}^{\dagger}-2),
\label{HKS}
\end{eqnarray}
where $E_l^\alpha $ is chromo-electric fields on the $l$ link, and it is
also generator of the gauge group. The mass spectrum can obtained directly
by solving the eigen-equation 
$H\Psi [U]=\epsilon _\Psi \Psi [U]$. 
Here $\epsilon _\Psi $ is the eigenvalue of $H$. When $a\to 0$, a huge
number of gauge configurations are correlated, and it is very difficult to
diagonalize the Hamiltonian with sufficient accuracy. For this reason, there
is no satisfactory (3+1)-D result in this formulation up to now. In a
feasible calculation, the gauge configuration space has to be truncated. An
inappropriate truncation scheme often violates the continuum limit, and
destroys the scaling behavior for the physical quantities. Therefore,
special care must be taken when choosing a truncation scheme.

On the other hand, the Hamiltonian method does have some advantages over the
Lagrangian method. It is relatively simple to obtain wave functions of the
hadronic states, and it is likely that more physical information on the
glueballs can be derived in this formulation.

\section{Brief Comments on Earlier Investigations}

We would like to mention here some earlier work on the Hamiltonian
formulation:

\noindent
{\bf a)} Strong coupling expansion with Pad\'e approximants \cite{Banks}.
Since the strong coupling expansion diverges at small $g$, some approximants
must be used to extrapolate the strong coupling series to the weak coupling
region. This introduces uncertainties and no conclusive results have been
obtained.

\noindent
{\bf b)} Eigen-equation method with truncated gauge configurations proposed
by Greensite \cite{Green1}. This is just the coupled cluster method with a
special truncation scheme. In \cite{Green1}, only the strong coupling region
was explored. However, the scaling region was not reached and no concrete
results were obtained.

\noindent
{\bf c)} Variational method. The vacuum energy $\epsilon_{\Omega}$ is
obtained by minimizing the expectation value of $H$ in the trial vacuum
state $\vert \Omega \rangle$, and the mass gap is obtained by minimizing the
expectation value of $H$ in the trial excited state $\vert \Psi \rangle$
with $\langle \Omega \vert \Psi \rangle=0$. Some earlier results \cite
{Arisue1} were consistent with asymptotic scaling predictions. However,
since the variation energies depend strongly on the choice of trial wave
functions, it is not clear whether this method can give reliable results for
the mass spectrum in a systematic way.

\noindent
{\bf d)} Models with exact ground state \cite{GZL}. Some modified
Hamiltonians with classical continuum limit the same as that of the
Kogut-Susskind (K-S) Hamiltonian were proposed. These models possesses an
exact ground state of the form $|\Omega \rangle =e^{S[U]}|0\rangle ,$ where $%
|0\rangle $ is the fluxless state and $S(U)$ is some gauge invariant
operator. The mass gaps were obtained by minimizing the excitation energies,
which was shown to have good scaling behavior in (2+1)-D theories. The
problem is that the modified Hamiltonian differs at quantum level from the
K-S Hamiltonian by a relevant operator and hence belongs to a universal
class different from the K-S Hamiltonian.

\noindent
{\bf e)} Linked cluster expansion method \cite{IPH}. While the Hamiltonian
can not be diagonalized exactly, its sub-matrix $\langle \Psi_i \vert H
\vert \Psi_j \rangle$ on a finite set of strong coupling basis $\Psi_i$ can
be diagonalized. This is a variant of strong coupling expansion, and some
approximants must be used to extrapolate the results into the weak coupling
region. However, such an extrapolation might lead to uncertainties.

\noindent
{\bf f)} Unitary transformation and variational method for LGT with fermions 
\cite{CL}. A variational form of hadronic wave functions that takes into
account the effect of sea quarks was proposed $|\Psi \rangle =e^{C{\bar \psi 
}\Gamma \psi }|0\rangle .$ Meson spectrum and chiral condensates were
obtained in low dimensional cases. They were consistent with scaling
predictions. This is a first step towards the more systematic coupled
cluster method.

To summarize, much efforts have been made in earlier investigations, but it
seems difficult to use them to study the continuum physics of a realistic
theory such as ${\rm {QCD}_4}$. More systematic and unambiguous methods are
required.

\section{Recent Advances: Coupled-Cluster Method}

In the 90's, one has seen considerable developments in Hamiltonian LGT for
approaching the scaling region. One of the most promising analytical method
is the coupled-cluster method with an appropriate truncation scheme \cite
{GCL}. We will discuss this method is some details.

\subsection{Coupled-cluster method (CCM)}

Let the vacuum state be

\begin{eqnarray}
\vert \Omega \rangle=e^{R[U]} \vert 0 \rangle,
\label{VAC}
\end{eqnarray}
The eigen-equation $H|\Omega \rangle =\epsilon _\Omega |\Omega \rangle $
becomes \cite{Green1,GCL,HZS}

\begin{eqnarray}
\sum_{l} \lbrace [E_l^{\alpha},[E_l^{\alpha},R]]+[E_l^{\alpha},R][E_l^{\alpha},R] \rbrace
- {2 \over g^4} \sum_{p} tr(U_p+U_{p}^{\dagger})
={2a \over g^2} \epsilon_{\Omega}.
\label{Eigen}
\end{eqnarray}
$R[U]$ is composed of various gauge invariant Wilson loops, and it can be
expanded in a series of graphs

\begin{eqnarray}
R[U]=\sum_{n} R_{n}[U]=\sum_{n,i} C_{n,i} \sum_x G_{n,i}[U].
\label{R}
\end{eqnarray}
For example, in (2+1)D SU(2) theory, $G_{1,1}=\Box $, $G_{2,1}=\Box \Box $, $%
G_{2,2}=\setlength{\unitlength}{.007in}%
\begin{picture}(75,13)(0,17)
  \put(10,10){\line(0,1){20}}
  \put(10,30){\line(1,0){40}}
  \put(50,30){\line(0,-1){20}}
  \put(50,10){\line(-1,0){40}}
\end{picture}
$,etc.

The eigen-equation (\ref{Eigen}) is a system of nonlinear equations for the
coefficients $C_{n,i}$. In practice, this equation must be truncated at some
finite $n$. Inappropriate truncation scheme would violate the continuum
limit of (13) and destroy the scaling behavior.

\subsection{The continuum limit of a graph}

The continuum limit of a generic graph $G_{n,i}$ has the form

\begin{eqnarray}
  G_{n,i}[U]=g^2 a^4[A_{n,i} ~ Tr ({\cal F}^2)
  +a^2 B_{n,i}  ~ Tr ({\cal D} {\cal F})^2+...].
  \label{CON}
\end{eqnarray}
For example, the elementary plaquette behaves as

\begin{eqnarray}
U_p= Re Tr (P e^{i \oint_\Box gA \cdot d x} )
=Re Tr [ 1 + i\oint_\Box gA\cdot d x - {1 \over 2} ( \oint_\Box
gA \cdot d x)^2 +... ].
\label{P}
\end{eqnarray}
Note that up to the $A^2$ term, it is not necessary for path to be ordering
because of the trace. Once the $A^2$ term is determined, we can supply the $%
A^3$ and $A^4$ terms to make up gauge invariant expressions. Thus with $x_0$
at its center,

\begin{eqnarray*}
\oint_\Box  A \cdot d x = \int_{-a/2}^{a/2} d x_\mu dx_\nu \left[\partial_\mu
A_\nu(x_0+x) - \partial_\nu A_\mu(x_0+x)\right] 
\end{eqnarray*}
\begin{eqnarray}
= a^2 {\cal F}_{\mu\nu} (x_0)  
+ {a^4 \over 24}({ \cal D}_{\mu}^2+{\cal D}_{\nu}^2) {\cal F}_{\mu\nu} (x_0)
+O(a^6,A^2).
\end{eqnarray}
Therefore, (\ref{P}) becomes

\begin{eqnarray}
Tr[U_p -1] = -
{ a^4 g^2 \over 2}  Tr ({\cal F}_{\mu\nu} {\cal F}_{\mu\nu})
+ {a^6 g^2 \over 24} Tr ({\cal D}_{\mu} {\cal F}_{\mu\nu})^2 +...
\end{eqnarray}
The continuum limit of other graphs can be calculated in a similar way.
Therefore, the operator $R$ has the continuum limit

\begin{eqnarray}
R[U]= - \int d^{D-1} x ~ [ \mu_0 Tr ({\cal F}_{\mu\nu} {\cal F}_{\mu\nu})
+ \mu_2 Tr ({\cal D}_{\mu} {\cal F}_{\mu\nu})^2 +...].
\end{eqnarray}

The vacuum state for the SU(2) gauge theory was first investigated by
Greensite \cite{Green2} and later by Arisue \cite{Arisue2} using simulation
method. Nice scaling behavior for $\mu_0$ and $\mu_2$ was obtained in the 3D
theory \cite{Arisue2}. These coefficients approach constant values in the
scaling region.

\subsection{Truncation scheme preserving the continuum limit}

The eigen-equation (\ref{Eigen}) has to be solved in some truncation
scheme. If in the expansion (\ref{R}), we let $n$ be the order of a graph ,
then we have

\begin{eqnarray}
\sum_{l} [E_l^{\alpha},[E_l^{\alpha},R_n]] 
\in R_{n} + lower ~ orders ~terms,
\end{eqnarray}
\begin{eqnarray}
\sum_{l}  [E_l^{\alpha},R][E_l^{\alpha},R] 
\in R_{n_1+n_2} +lower ~ order ~ terms.
\label{Trun}
\end{eqnarray}
Thus the last term must be truncated such that the orders of graphs
appearing in (\ref{Eigen}) do not exceed some finite order $N$. A
conventional prescription \cite{Green1,LW} was that in (\ref{Trun}), only
graphs with order $N$ were preserved, while graphs with order greater than $%
N $ were discarded.

We \cite{GCL} have shown that these truncation schemes violate the continuum
limit of the term (\ref{Trun}), and hence destroy the scaling behavior of $%
\mu _0$ and $\mu _2$. The essential point is that, since a general graph $%
G_i $ has the continuum limit(\ref{CON}), the continuum limit of $%
[E_l^\alpha ,G_i][E_l^\alpha ,G_j]$ is

\begin{eqnarray}
[E_l^{\alpha},G_i][E_l^{\alpha},G_j]  \propto g^2 a^6 ~Tr({\cal D} {\cal F}_{\mu,\nu})^2+...
\end{eqnarray}
For this equation to be valid, all graphs generated from the l.h.s. must be
included. On the contrary, if some graphs were kept and others were
discarded, the continuum limit would change into $a^4g^2Tr({\cal F}_{\mu \nu
}{\cal F}_{\mu \nu })$, leading to seriously wrong continuum behavior.

Based on this consideration, we proposed a truncation scheme \cite{GCL} that
respects the continuum limit,

\begin{eqnarray*}
\sum_{l} \lbrace [E_l^{\alpha},[E_l^{\alpha},\sum_{n}^{N} R_{n}(U)]]
+\sum_{n_1+n_2 \le N}[E_l^{\alpha},R_{n_1}(U)][E_l^{\alpha},R_{n_2}(U)] 
\rbrace
\end{eqnarray*}
\begin{eqnarray}
- {2 \over g^4} \sum_{p} tr(U_p+U_{p}^{\dagger})
={2a \over g^2} \epsilon_{\Omega}.
\label{b2}
\end{eqnarray}

To confirm the validity of this method, we must test the convergency of the
truncation scheme at large $N$ and check the scaling behavior of various
quantities, in particular, the coefficients $\mu _0$, and $\mu _2$ of the
vacuum wave function and the glueball masses $m(J^{pc})$.

\subsection{Results}

We have examined this method for several lattice field systems:

\noindent
{\bf a)} (2+1)D U(1) theory \cite{FLG}. $\mu _0$ and the mass gap $m_A$ are
calculated up to 8th order. The results show clear tendency of convergence
and exponential scaling. The mass gap agrees with MC and other analytic
results. (See Fig. 1).

\noindent
{\bf b)} (2+1)D SU(2) theory \cite{GCL,Chen}. Reasonable results with power
scaling for $\mu _0$, $\mu _2$ and $m_S$ are obtained even at the 3rd order.
Convergent results are obtained up to $7th$ order. The continuum limit of $%
m_S\approx 1.4~e^2$ is somewhat lower than Teper's MC results. (See Fig. 2).

\noindent
{\bf c)} (2+1)D SU(3) theory \cite{CLG}. Up to the 3rd order, we obtained $%
\mu _0$, $\mu _2$, $m(0^{++})$ and $m(0^{--})$, consistent with power
scaling, $m(0^{++})\approx 2.09e^2,$ and $m(0^{--})\approx 3.71e^2$. (See
Figs. 3, 4).

\noindent
{\bf d)} (1+1)D O(3) $\sigma $ model \cite{FGL}. Up to the 8th order, the
triplet mass $m_T$ is consistent with MC results, indicating that asymptotic
scaling occurs at rather large values of $1/g^2$. At smaller values of $%
\beta $, there exists a scaling region according to the whole $\beta $
function. We confirm that $\mu _0m_T\propto const.$ in this region. (See
Fig. 5).

\noindent
{\bf e)} (1+1)D O(2) model \cite{LFG} (Hamiltonian formulation of 2D XY
model). Convergency was examined up to the 10th order. There is a clear
signal of KT phase transition at $g^2\approx 1$. The critical point and
critical exponent are consistent with results by other methods. (See Fig. 6).

\noindent
{\bf f)} Estimates of ${\rm {QCD}_4}$ glueball masses from ${\rm {QCD}_3}$
results by dimensional reduction \cite{MPL}. In the strong coupling limit or
large $N_c$ limit, the fixed time vacuum expectation value of an operator $%
O(U)$ in D dimensional confinement theory ($2<D\le 4$) correponds to the
path integral expression $<O(U)>$ in (D-1)-dimensional theory. For long
wavelength excitations, the mass ratios should be approximately the same for
(3+1)D and (2+1)D. Combining our updated ${\rm {QCD}_3}$ data $%
m(0^{++})\approx (2.15\pm 0.06)e^2$, and the recent MC data for the ${\rm {%
QCD}_3}$ string tension $\sigma =(0.554\pm 0.004)e^2$, we obtain

\begin{eqnarray}
{m(0^{++}) \over {\sqrt \sigma}} \approx 3.88 \pm 0.11,
\end{eqnarray}
which is to be compared with the recent MC data for ${\rm {QCD}_4}$: $%
m(0^{++}) / {\sqrt \sigma} \approx 3.95$. We see that the dimensional
reduction relation works quite well in this case. If we further use the MC
result for the ${\rm {QCD}_4}$ string tension $\sigma \approx 0.44 ~GeV$, we
obtain for ${\rm {QCD}_4}$:

\begin{eqnarray}
m(0^{++})  \approx 1.71 \pm 0.05 ~Gev,
\end{eqnarray}
which is consistent with the IBM data $M(0^{++})=1.740 \pm 0.071$ \cite{IBM}.

\noindent
{\bf g)} Preliminary calculations of QCD spectrum in (3+1)D \cite{Luo}. We
have also calculated the $0^{++}$, $0^{--}$ and $1^{+-}$ glueball masses in $%
{\rm {QCD}_4}$ in low order approximations. Although the scaling behavior
for these glueball masses has not been achieved, we do observe plateaus for
glueball mass ratios in $\beta \in (6.0,6.4)$. From the plateaus we obtain

\begin{eqnarray}
{M(0^{--}) \over M(0^{++})}=2.44 \pm 0.05 \pm 0.20, ~~~
{M(1^{+-}) \over M(0^{++})}=1.91 \pm 0.05 \pm 0.12,
\label{data}
\end{eqnarray}
where the first error is the error of the data in the plateau, the second
error is the estimated error due to finite order truncation. Our results are
in good agreement with the MC and t-expansion results.

\section{Other Developments}

\noindent
{\bf a)} Improved Hamiltonians. The idea of improved lattice actions was
proposed in the early 80's \cite{Sym}. Recently the improved actions have
attracted much attention \cite{Lepage}. The purpose is to push the $O(a)$
errors to higher orders and remove the tadpole lattice artifacts. Using
improved actions, one can obtain results on coarser lattice, which was
previously obtained on finer lattice. The computation time may then be
greatly reduced. Starting from Lepage's improved Lagrangian, an improved
Hamiltonian \cite{Imp} can be straightforwardly derived using the transfer
matrix method or the Legendre transform method. However, the color electric
energy becomes an infinite series with long range terms. This deficiency can
be cured by the Legendre transformation of a suitable improved Lagrangian
with infinite time-like terms, yielding the same order of improvement but
with only local and nearest neighbor interactions \cite{Imp}.

\noindent
{\bf b)} Universality of LGT in Hamiltonian and Lagrangian formulations \cite
{Hamer}. Hamer et al. recently derived the relation between the coupling
constants $g_H$ and $g_L$ in both formulations. Relations between velocity
of light in both formulations were also derived. Using these relations, a
comparison was made for (2+1)D SU(2) theory between various Hamiltonian
calculations and MC results in the Lagrangian formulation. A striking
demonstration of universality between both formulation was obtained.

\noindent
{\bf c)} Calculation of hadronic structure functions \cite{HN}. A major
difficulty in Hamiltonian LGT is that a great number of correlated
configurations are involved. Recently Kr\"oger and Scheu showed in a scalar
model that by using specific reference frame (e.g. Breit frame), the number
of relevant configurations may be greatly reduced, so that they could obtain
nice results for distribution function consistent with the theoretical
prediction.

\section{Conclusions and Outlook}

Recent work on Hamiltonian LGT shows that with appropriate calculation
schemes (CCM with continuum limit preserved truncation, Breit frame, etc.),
we can efficiently enter the scaling region and obtain reliable results.
More calculations on (3+1)D should be performed for further establishing the
advantage of the Hamiltonian formulation in spectrum and wave function
calculations. As the number of correlated, configurations increase
considerably from (2+1)D to (3+1)D, improved Hamiltonians may be necessary
for efficient calculations in (3+1)D.

We would like to thank Q. Chen, X. Fang. H. Kr\"oger, J.M. Liu and D.
Sch\"utte for collaboration, C. Hamer, J.J. Liu and N. Scheu for
discussions. This work was support by National Natural Science Foundation
under grant numbers 19575075, 19605009 and 19677205, and by National
Education Committee under grant number jiao-wai-si-liu [1996] 644.

\end{document}